\documentclass[preprint2]{aastex}

\slugcomment{Not to appear in Nonlearned J., 45.}

\shorttitle{Reflection effect in the interacting binaries and extrasolar
planets}
\shortauthors{J. Budaj}

\begin{document}


\title{
A simple model of the reflection effect for the interacting
binaries and extrasolar planets
}


\author{J. Budaj}
\affil{Astronomical Institute, Slovak Academy of Sciences,
Tatransk\'{a} Lomnica 05960, Slovak Republic}
\email{budaj@ta3.sk}


%\altaffiltext{1}{Visiting Astronomer, Cerro Tololo Inter-American Observatory.
%CTIO is operated by AURA, Inc.\ under contract to the National Science


\begin{abstract}

Extrasolar planets are a natural extension of the interacting binaries 
towards the companions with very small masses and similar tools might 
be used to study them. Unfortunately, the generally accepted treatment 
of the reflection effect in interacting binaries is not very suitable
to study cold objects irradiated by hot objects or extrasolar planets.

The aim of this paper is to develop a simple model of the reflection
effect which could be easily incorporated into the present codes
for modeling of interacting binaries so that they can be used to study
above mentioned objects.

Our simple model of the reflection effect takes into account
the reflection (scattering), heating and heat redistribution over
the surface of the irradiated object. The shape of the objects is 
described by the Roche potential and limb and gravity darkening
can be taken into account. The orbital revolution and rotation
of the planet with proper Doppler shifts for the scattered and thermal
radiation are also accounted for.
Subsequently, light-curves and/or spectra of exoplanets were
modeled and the effects of the heat redistribution, 
limb darkening/brightening, (non-)grey albedo, and non-spherical shape 
were studied. Recent observations of HD189733b, WASP12b, and Wasp-19b 
were reproduced reasonably well. HD189733b has low Bond albedo
and intense heat redistribution. Wasp-19b has low Bond albedo
and low heat redistribution.

We also calculate the exact Roche shapes and temperature
distribution over the surface of all 78  transiting 
extrasolar planets known so far. It is found that the departures
from the sphere vary considerably within the sample. 
Departures of about 1\% are common. 
In some cases (WASP-12b, WASP-19b, WASP-33b) 
departures can reach about 14, 12, and 8\%, respectively. 
The mean temperatures of these planets 
also vary considerably from 300 K to 2600 K. The extreme cases are 
WASP-33b, WASP-12b, and WASP-18b with mean temperatures 
of about 2600, 2430, and 2330 K, respectively.

\end{abstract}


\section{Introduction}

There are sophisticated computer codes for calculating and inverting 
light curves or spectra of binary stars with various shapes or 
geometry including the Roche model 
\citep{lucy68,wd71,wood71,md72,rucinski73,hill79,pe81,zhang86,
djurasevic92,drechsel94,lh96,hadrava97,oh00,bs02,pribulla04,
pavlovski06,tamuz06}.
The Wilson \& Devinney (WD) code is most often used and is continuously
being improved or modified \citep{kallrath98,pz05}.
The reflection effect is taken into account in most of these codes.

The standard description of this effect is given in \citet{wilson90}. 
This effect is understood in the following way:
A surface element of star A is irradiated by many surface elements
of star B. A fraction of impinging energy called bolometric albedo
\citep{rucinski69} is converted into the heat which rises the local 
temperature and re-radiates the energy on the day side 
of the object. Rest of the impinging energy is plunged into 
the star. Typically, bolometric albedo is set to 1 for radiative 
and 0.5 for convective envelopes. An increase of the temperature on 
the day side of one object triggers a secondary reflection effect 
on the second object and one, two or several iterations are 
allowed to converge to a final state. Roche model, limb darkening 
and gravity darkening are taken into account. This model does 
a good job for many interacting binaries.
%\footnote{
%One might point out that the name "reflection effect" is misleading
%in this context since the effect as it is described above has nothing
%to do with the reflection (process of light scattering) and it deals 
%only with the heating and re-radiating.
%}
%It might be interesting to note that there are casses 
%when one needs to addopt strange values of some parameters like
%extremely large gravity darkening coeficients to cool down the vicinity
%of the substelar (L1) point of the iradiated object since the model
%predicts very hot temperatures in this region as recently pointed out 
%by Desmet et al. (2009).

However, I argue that the above mentioned model of the reflection 
effect should be revisited. There has been a lot of progress in 
the field since the time this reflection effect was developed.
New types of very cool objects such as brown dwarfs and extrasolar 
planets have been discovered. These new areas evolve rapidly and 
produce interesting results. In many respects, an extrasolar planet 
can be understood as an interacting binary with a small mass companion.
It is thus an attractive idea to apply these new results from 
the extrasolar planets to interacting binaries and vice versa. 
For example, models of extrasolar planets can provide day-night heat 
redistribution and reflected light while models of interacting 
binaries can provide sophisticated Roche geometry.

The standard model of the reflection effect faces several problems 
which prevent its application to cool objects irradiated by hot 
objects and to extrasolar planets.
When it comes to very cold strongly irradiated objects
a considerable amount of energy might be reflected off the surface.
The above mentioned reflection effect in the interacting binaries
neglects this reflected light which can be essential,
especially at the shorter wavelength.
Well known example of such effect are the planets and moons in our 
Solar system in the visible light. This reflected light
bears the spectroscopic signatures of the hot irradiating star and
is not converted into the heat and re-radiated.
It is commonly taken into account in models of hot-Jupiters or 
planets in general \citep{seager00,sudarsky05}.
The definition of the albedo in the planetary sciences is 
almost the opposite of its meaning in the interacting binaries. 
In planets, the albedo is a fraction of the impinging 
energy which is reflected off the object and is not absorbed and 
converted into the heat.
Reflected and re-radiated photons may also have completely 
different Doppler shifts depending on the mutual velocities 
of the two objects and observer.

Moreover, some portion of the energy absorbed on the day side
can be transferred to and irradiated from the night side.
Various authors developed different parametrization of this
effect in connection with the extrasolar planets 
\citep{guillot96,burrows06,ca10}.
Calculations of atmosphere models of extrasolar planets
\citep{hubeny03,barman05,burrows08,fortney08} demonstrate
that there is a deep temperature plateau which turns off 
the convection in the atmospheres. Convection and vertical energy 
transport operates only at very deep layers. Observations of 
hot-Jupiters indicate that only a very small fraction of 
the impinging radiation ($10^{-4}$) could have been plunged into 
the object \citep{burrows07}. Hydrodynamical simulations 
\citep{dl08,showman09,mr09} of the atmospheres of extrasolar 
planets indicate that there are very strong horizontal currents 
and jets which can effectively redistribute and circulate 
the energy between the day and night side of the planet especially 
along the lines of constant latitudes. How much energy gets 
redistributed to the night side depends mainly on the complex 
structure and dynamics of the surface layers.

The above mentioned effects and findings must be taken into 
account in the new or revisited model of the reflection effect 
when modeling very cold components of interacting binaries and 
extrasolar planets. At the same time, many transiting exoplanets 
are so close to their host stars that their shape may depart 
from the sphere and be best described by the Roche model.
  
The main purpose of this paper is to develop a new simple
model of the reflection effect which would consider
the reflected (scattered) light, heating (absorption of the light),
and heat redistribution over the surface and which could be used 
to model the cold components of interacting binaries and extrasolar 
planets. The model was included into our code {\sc{shellspec}} 
\citep{br04,budaj05}. This program  was originally designed 
to calculate light-curves, spectra and images of interacting 
binaries immersed in a moving circumstellar environment which 
is optically thin. The code is freely available at
{\tt http://www.ta3.sk/$\sim$budaj/shellspec.html}
and might be used to study various effects observed or expected in 
the interacting binaries or extrasolar planets.
A Fortran90 version of the code which also solves the inverse problem
was created by \citet{tkachenko10}.

\section{Roche model}
\label{rg}
In the {\sc{shellspec}} code the Roche model serves as a boundary 
condition for the radiative transfer in the circumstellar matter.
Both objects, star and companion, may have shapes according to 
the Roche model for detached or contact systems. Descriptions of 
the Roche model can be found in \citet{kopal59}, \citet{pk64},
and many other papers and books.
Let us assume a Cartesian coordinate system (x,y,z) centered on one 
of the stars (labeled as 1) such that the companion (labeled as 2) 
is at (1,0,0) and revolves around the z axis in the direction of 
positive y axis. Let the mass ratio, $q$, always be $m_{2}/m_{1}$ 
or `companion/star' and $q<1$ will indicate the companion is 
lighter while $q>1$ means the central star is lighter.
Then, the normalized Roche potential, C, is expressed as:
%\begin{equation}                                     
\begin{eqnarray}
%\begin{array}
C(x,y,z)=   \nonumber  \\
\frac{2}{(1+q)r_{1}}+\frac{2q}{(1+q)r_{2}}+
\left(x-\frac{q}{1+q} \right)^{2}+y^{2}
%\end{array}
\end{eqnarray}
%\end {equation}
where 
\begin{equation}
%\begin{eqnarray}
\begin{array}
rr_{1}=\sqrt{x^{2}+y^{2}+z^{2}}   \\\\
r_{2}=\sqrt{(x-1)^{2}+y^{2}+z^{2}}.
\end{array}
%\end{eqnarray}
\end{equation}
The Roche surface of a detached component is defined as 
an equipotential surface $C_{s}=C(x_{s},y_{s},z_{s})$ passing 
through the sub-stellar point $(x_{s},y_{s},z_{s})$ (point on 
the surface of the star in between the stars, 
$0<x_{s}<1, y_{s}=z_{s}=0$) which is localized by the `fill-in' 
parameter $f_{i}\leq 1$. We define this by:
\begin {equation} 
f_{i}=x_{s}/L_{1x}, ~~~~~~~f_{i}=(1-x_{s})/(1-L_{1x})
\end {equation}
for the primary and the secondary, respectively.
$L_{1x}$ is the x coordinate of the L1 point L1($L_{1x},0,0$).

Derivatives of the Roche potential are:
\begin {equation}
C_{x}=\frac{\partial C}{\partial x}~,~~~
C_{y}=\frac{\partial C}{\partial y}~,~~~ 
C_{z}=\frac{\partial C}{\partial z}
\end {equation}  

Gravity darkening of the non spherical objects
is taken into account by varying the surface 
temperature according to the von Zeipel's law:
\begin {equation}
T/T_{p}=(g/g_{p})^{\beta}
\label{tgd}
\end {equation}  
where $g$ is the normalized surface gravity, $\beta$ is the gravity 
darkening coefficient, 
$T_{p},g_{p}$ are the temperature and gravity at the rotation pole.
The normalized gravity vector is ${\bf g}=(C_{x},C_{y},C_{z})$ and:
\begin {equation}
g=\sqrt{C_{x}^{2}+C_{y}^{2}+C_{z}^{2}}.
\end {equation}  
%The gravity darkening factor of the surface intensity is then 
%calculated as:
%\begin {equation}                                    		
%f_{GD}=B_{\nu}(T)/B_{\nu}(T_{p}).
%\end {equation}  
Notice, that there is an imminent singularity in the calculations
in the vicinity of L1, L2 points since gravity falls to zero 
which drags temperatures (a denominator in many equations) to zero. 
We avoid the problem by setting the lowest possible value of 
$g/g_{p}=10^{-4}$.

Limb darkening is taken into account using the quadratic limb 
darkening law:
\begin {equation}
I(\theta)=I(0)f_{LD}
\label{ld}
\end {equation}
\begin {equation}
f_{LD}=1-u_{1}(1- \cos \theta)-u_{2}(1- \cos \theta)^{2}
\end {equation}
and by calculating the cosine of the angle $\theta$ between 
the line of sight unit vector ${\bf n}=(n_{x},n_{y},n_{z})$ 
and a normal to the surface:
\begin {equation}
\cos \theta=-{\bf n}.{\bf g}/g
=-\frac{n_{x}C_{x}+n_{y}C_{y}+n_{z}C_{z}}
{\sqrt{C_{x}^{2}+C_{y}^{2}+C_{z}^{2}}}.
\end {equation}

\section{Irradiation and heat redistribution}

In this section we will describe our treatment of the mutual irradiation 
of the objects. It can be applied to both objects but
we will neglect multiple reflections between the two objects since
this is not essential if one of them is much less luminous.
We will distinguish between three separate processes:
{\bf reflection} of the light off the object (or scattering which does not 
produce any heating of the irradiated surface); {\bf heating} of 
the irradiated surface (day side) by the absorbed light; and subsequent
{\bf heat redistribution} over the entire surface of the object.
Let's assume that the day side of a planet is irradiated by the star then
the impinging flux at a location ${\bf r}$ from star at ${\bf r_*}$ is:
\begin {equation}
F_{ir}= \cos \delta \frac{R_{\star}^{2}}{({\bf r}-{\bf r_*})^{2}}
\sigma T_{\star}^4
\label{fir}
\end {equation}
where $R_{\star},T_{\star}$ are radius and effective temperature of the
star, $\delta$ is an irradiating angle which is the zenith distance 
of the center of the star as seen from the surface of the planet:
%\begin {equation}
\begin {eqnarray}
\cos \delta=\frac{({\bf r}-{\bf r_*}).{\bf g}}{|{\bf r}-{\bf r_*}|g}=
\nonumber \\
\frac{(r_{x}-r_{x*})C_{x}+(r_{y}-r_{y*})C_{y}+(r_{z}-r_{z*})C_{z}}
{|{\bf r}-{\bf r_*}| 
\sqrt{C_{x}^{2}+C_{y}^{2}+C_{z}^{2}}}.
\end {eqnarray}
%\end {equation} 
Now, let's define two local parameters 
$A_{B}(\alpha,\beta)$, $P_{r}(\alpha,\beta)$ on the day side of 
the planet (irradiated side of an object) where
$\alpha, \beta$ are the longitude and latitude, respectively,
measured from the sub-stellar point.
$P_{r}$ will be the local heat redistribution parameter and
$A_{B}$ will be the local Bond albedo of the surface.
Consequently $A_{B}F_{ir}$ is flux immediately
reflected off the surface, $(1-A_{B})F_{ir}$ will be a fraction of 
the irradiating energy which is converted into the heat, 
$P_{r}(1-A_{B})F_{ir}$ will be a part of the latter which is 
redistributed over the day and night side of the object,
while the remaining part, $(1-P_{r})(1-A_{B})F_{ir}$, will heat 
the local area.
\footnote{Note that our local $P_{r}$ should not be confused with 
the global $P_{n}$ parameter of \citet{burrows06}. $P_{n}$ is a 
fraction of the impinging stellar radiation which is transferred to 
and reradiated from the night side, $P_{n}\approx P_{r}(1-A_{B})/2$.
$P_{n}$ is from the interval 0-0.5 while $P_{r}$ runs from 0 to 1.
$P_{r}$ is a direct indicator of the heat redistribution while 
the $P_{n}$ parameter only reflects a combination of the heat 
redistribution and Bond albedo. 
%??Small values of $P_{r}$ always mean that 
%the heat redistribution is not effective while small values of 
%$P_{n}$ do not always mean that the heat redistribution is not 
%effective.
}
Then the energy conservation
for the day-night heat transport can be written as:
\begin {equation}
\int\!\!\!\!\int P_{r}(1-A_{B})F_{ir}dS_{day}=
\int\!\!\!\!\int \sigma T^{4}_{dn}dS_{day+night}
\label{dn}
\end {equation}
Let's assume, of simplicity (or in the absence of a better approximation), 
that $A_{B}(\alpha,\beta),P_{r}(\alpha,\beta)$ are constant
and the heat is homogeneously redistributed over the day and night sides
so that the surface has a constant temperature $T_{0}$. Then,
\begin {equation}
P_{r}(1-A_{B})\int\!\!\!\!\int F_{ir}dS_{day}
/\int\!\!\!\!\int dS_{day+night}=\sigma T_{0}^4.
\label{dn2}
\end {equation}
In case the planet has a spherical shape and is far from the star 
this reduces to:
\begin {equation}
T_{0}^4=\frac{1}{4}P_{r}(1-A_{B})\frac{R_{\star}^{2}}{d^{2}}T_{\star}^{4}.
\end {equation}
Let's explore another case, namely that the horizontal circulation 
on the planet along the lines of constant latitude is so strong that 
it dominates the day-night heat transport and the equilibrium surface
temperature, $T_{1}$, will be a function of latitude.
In this case and the assumptions above (spherical planet far from the star),
one can consider an energy conservation for a fixed latitude:
\begin {equation}
P_{r}(1-A_{B})\int F_{ir}\cos(\beta)d\alpha=
\sigma T_{1}^4(\beta)\int\cos(\beta)d\alpha
\end {equation}
where
\begin {equation}
\cos(\delta)=\cos(\alpha)\cos(\beta).
\end {equation}
The solution is that the temperature depends on the fourth root of
$\cos(\beta)$:
\begin {equation}
T_{1}^4(\beta)=\frac{1}{\pi}
P_{r}(1-A_{B})\frac{R_{\star}^{2}}{d^{2}}T_{\star}^{4}\cos(\beta)=
\frac{4}{\pi}T_{0}^{4}\cos(\beta).
\end {equation}

The study of these two extreme cases lead us to suggest a heat
redistribution model in which the day-night heat transport is 
a linear combination of the two cases mentioned above. 
Namely, we will express the surface temperature 
in the following way:
\begin {equation}
T^4_{dn}(\beta)=T_{0}^{4}[P_{a}+P_{b}\cos(\beta)]
\label{dnmodel}
\end {equation}
where $P_{a}, P_{b}$ are the 'zonal temperature redistribution 
parameters'. $P_{a}=<0,1>$ and $P_{b}$ is to be determined 
from Eq.\ref{dn} so that the total energy budget is conserved. 
From Eq. \ref{dn},\ref{dn2} and \ref{dnmodel} we obtain:
\begin {equation}   
P_{b}=(1-P_{a}) \frac{\int\!\!\!\!\int dS_{day+night}}
{\int\!\!\!\!\int \cos \beta dS_{day+night}}.
\end {equation}   
It can be shown that in case of a spherical planet far from the star
\begin {equation}
P_{b}=\frac{4}{\pi}(1-P_{a})
\end {equation}
and $P_{b}=<0,4/\pi>$.
Notice, that $P_{a}$ is a measure of the effectiveness of the homogeneous
temperature distribution over the surface versus the zonal distribution.
It is intimately linked with the effectiveness of the heat flows
along the meridians versus parallels.
 
Finally, the temperature distribution on the surface of the irradiated
planet will be:
\begin {equation}
T^4=T_{ir}^{4}+T_{dn}^{4}+T_{old}^{4}
\label{irmodel}
\end {equation}
where $T_{ir}^{4}=(1-P_{r})(1-A_{B})F_{ir}/\sigma$ on the day side,
$T_{ir}^{4}=0$ on the night side, and $T_{old}$ is the prior temperature
distribution over the surface in the absence of the irradiation 
(including the gravity darkening etc.).
It should be noted that imposing the external irradiation on one side 
of the object can alter the original temperature distribution.
\citet{budaj09} argue that the core cooling rates ($T_{old}$) from 
the day and night
side of a strongly irradiated planet may not be the same and
that the difference depends on several important parameters, such as the
effectiveness and the depth where the day-night heat transport occurs,
the stellar irradiation flux, and vertical redistribution of the opacities,
atmospheric abundances, and/or presence of the stratospheres.

Once we know the temperature distribution over the surface,
one can approximate the mo\-no\-chro\-ma\-tic flux from 
the surface as being composed of two parts:
\begin {equation}
F_{\nu}=F_{\nu}^{reflect}+F_{\nu}^{thermal}
\end {equation} 
Reflection depends on the surface albedo $A_{\nu}$ and has to take into 
account the mutual velocities:
\begin {equation}
F_{\nu}^{reflect}=A_{\nu_{2}}F_{\nu_{1}ir}           
\end {equation}
\begin {equation}
F_{\nu_{1}ir}=\cos \delta \frac{R_{\star}^{2}}{({\bf r}-{\bf r_*})^{2}}
F_{\nu_1}^{\star}
\end {equation}
where $F_{\nu_1}^{\star}$ is properly shifted flux emerging from 
the surface of the irradiating star. The Doppler shifts are 
the following 
\begin {equation}
\nu_{2}=\nu \left( 1- \frac{v_{z}}{c} \right)
\end {equation}
\begin {equation}
\nu_{1}=\nu \left( 1- \frac{v_{z}+v_{2}}{c} \right)
\end {equation}
\begin {equation}
v_{2}=-\frac{({\bf r}-{\bf r}_{\star}).({\bf v}-{\bf v}_{\star})}
{|{\bf r}-{\bf r}_{\star}|}
\end {equation}
where ${\bf v}$ is the velocity field vector at the given point on the
irradiated surface specified by the vector ${\bf r}$, ${\bf v}_{\star}$ 
is the velocity of the center of mass of the irradiating star, and
$z$ coordinate points to the observer.
\footnote{Note that generally Doppler shifts in the scattered and thermal
radiation are not the same and may not be trivial. 
More detailed treatement would require high resolution radiative 
transfer in the irradiated atmospheres for a set of radial 
planet-star velocities.}
To calculate the reflected intensity we assume that the reflection is
isotropic in which case:
\begin {equation}
I_{\nu}^{reflect}=F_{\nu}^{reflect}/\pi.
\end {equation} 
Finally, $F_{\nu}^{thermal}$ can be approximated by the flux emerging
from the non-irradiated model atmosphere with the effective temperature
equal to the surface temperature of the irradiated object given by
Eq.\ref{irmodel}.
\begin {equation}
F_{\nu}^{thermal}=F_{\nu_{2}}(T_{eff}=T)
\end {equation} 
and the associated intensity is given by:
\begin {equation}
I_{\nu}^{thermal}=I_{\nu}(0)^{thermal}f_{LD}
\end {equation}
\begin {equation}
I_{\nu}(0)^{thermal}= \frac{F_{\nu}^{thermal}}{\pi (1-u_{1}/3-u_{2}/6)}
\end {equation}
In this way we fully include into account the mutual velocities of 
the objects and observer, the rotation of the reflecting object,
its limb darkening but neglect the rotation of the irradiating object. 
In some cases the later can be easily taken into account by feeding 
the code with the precalculated rotationally broadened spectrum 
$F_{\nu}^{\star}$.

Local Bond albedo used here is a weighted monochromatic albedo $A_{\nu}$:
\begin {equation}
A_{B}=\frac{\int A_{\nu}F_{\nu}^{\star}d\nu}{\int F_{\nu}^{\star}d\nu}.
\end {equation} 
One has to keep in mind that our albedo refers
to the reflected light only (not to the absorbed and re-radiated light).
In case the irradiated object is very cold compared to the irradiating
object there is a clear distinction between its thermal radiation
and reflected radiation. However, if the two objects have comparable
temperatures it is almost impossible to distinguish whether a particular 
photon was scattered or absorbed and re-radiated. In this case it is still
possible to use our formalism and e.g. approximate the albedo by the single
scattering albedo to cope with the problem.

It might be convenient to define the mean temperature of the whole 
distorted object:
\begin {equation}
T_{mean}^{4} \equiv \frac{\int T^{4}dS}{\int dS}.
\end {equation} 
Note that it does not takes into account the reflected light
and should not be confused with the effective temperature
or brightness temperature.

\section{Application to extrasolar planets}

Extrasolar planets are an attractive test-bed for such models.
They are a natural extension of the interacting binaries
towards cool and small mass companions. Their radiation is governed
by the stellar insolation. Very sophisticated atmosphere models
of exoplanets are being developed 
\citep{seager00,hubeny03,barman05,burrows08,fortney08}.
At the same time, models with various simplifications
are often very useful to study various effects 
\citep{hansen08,kt09,ms09,ca10}.

\subsection{Shapes of the transiting extrasolar planets}

Our knowledge about the size of exoplanets comes mainly
from the observations of transits. A majority exoplanet studies
assume that the exoplanets have a spherical shape.
The precision of the photometric measurements is advancing 
revealing more and more details. Recently, \citet{kt09} attempted
to model the light curve and transit assuming the spherical shape
but taking into account the intrinsic night side emission of 
the planet during the transit. They found that in the infrared
region the effect is of the order of $10^{-4}$.
\citet{sh02}, \citet{bf03},\citet{cw10} assumed the shape of 
the rotational ellipsoid.
\citet{welsh10} used the Roche potential approximation to describe
the shape of the parent star and to explain the ellipsoidal variations
of HAT-P-7.
Nevertheless, many transiting exoplanets are so close to their host
stars that their shape is best described by the Roche model.
Very recently, \citet{li10} developed a very sophisticated model
for the tidally distorted exoplanet WASP-12b.

Our code was applied to transiting extrasolar planets and exact 
Roche model shape of all the transiting exoplanets was calculated. 
Circular orbit with the radius equal to the semi-major axis and 
synchronized rotation of the planet was assumed.
The results are displayed in the Table 1.
It lists $R_{sub}$, $R_{back}$, $R_{pole}$, $R_{side}$, $R_{eff}$ 
which are the radii at the sub-stellar point, anti-stellar point, 
rotation pole, on the side, and the effective radius of the planet, 
i.e. the radius of the sphere with the same volume as the Roche
surface, respectively. Departures from the sphere 
are measured by the value of the $R_{sub}/R_{pole}$.
It was also assumed that the observed radius of the planet is 
equal to $R_{side}$ which has almost no effect on the relative
proportions of most of the planets but may slightly 
underestimate the departures from the sphere for a few highly 
distorted planets.
%It was also assumed that the measured radius of the planet is equal
%to its side radius.

One can see that the departures from the spherical shape vary by 
several orders of magnitude. Departures of about 1\% are common.
About 8\% of planets (all of them have semi-major axis smaller 
than 0.03 AU) have departures larger than 3\%.  These include: 
OGLE-TR-56b, WASP-4b, and CoRoT-1b. The extreme cases are 
WASP-12b, WASP-19b, and WASP-33b when the departures can exceed 
14\%, 12\% and 8\%, respectively. 
This is clearly comparable to the precision of the planet
radius determination or the transit radius effect.
Moreover, this shape distortion would be even higher 
if one was to consider the eccentricity of the orbit
and assume the periastron distance instead of the semi-major axis.
%The shape distorsion might also be slightly higher if one was 
%to assume that the measured radius of the planet is in fact 
%an effective cross-section of the Roche shape.
The highest departures seems to be along the line joining the objects
(the $R_{sub}/R_{pole}$ parameter) which would affect mainly 
the overall light-curve. During the transit
event, it is mainly $R_{side}/R_{pole}$ parameter which is most
relevant and this does not acquire such high values (about 2\%
in case of WASP-12b, WASP-19b, and WASP-33b).

The observed radius of the planet should be associated with 
the cross-section of the planet during the transit which is 
represented by $R_{side}, R_{pole}$.
On the other hand, theoretical radius, e.g. from the evolutionary
calculations, might be associated rather with the effective radius 
of the planet.
One has to keep the findings above in mind when interpreting 
the planet radius measurements or calculations 
\citep{gs02,burrows07,fortney07,baraffe08,leconte10}. 

Apart from the shape of these planets, I calculated also 
a temperature distribution over the surface for all these exoplanets.
This is rather cumbersome to tabulate and the result depends 
heavily on the free parameters of the model that is why I provide 
the reader with the code which can do that. In the Table 1 
only the mean temperature of each transiting planet is listed.
This was calculated assuming its Roche shape and
$A_{B}=0.1, P_{r}=0.5, P_{a}=0.5, T_{old}=100K$.
These mean temperatures vary considerably from 300 to 2600K.
The hottest planets are WASP-33b, WASP-12b, and WASP-18b with 
the mean temperatures of about 2600, 2430, and 2330 K, 
respectively. The separation of these extremely hot planets
from the host star is smaller than 0.03 AU and temperatures 
of planets at a wider separation decrease accordingly.

\subsection{Light curves of the extrasolar planets}

In this section, we apply our model to the light-curves of 
extrasolar planets HD189733b and Wasp-12b.
We will start with HD189733b. The planet properties were 
determined by \citet{winn07}. \citet{knutson07} obtained 
superb light curve of the planet which covers more than half 
of the orbit. It was observed at 8 micron in the infrared region.

\begin{figure}
\centerline{\includegraphics[width=6.cm,angle=-90,clip=]{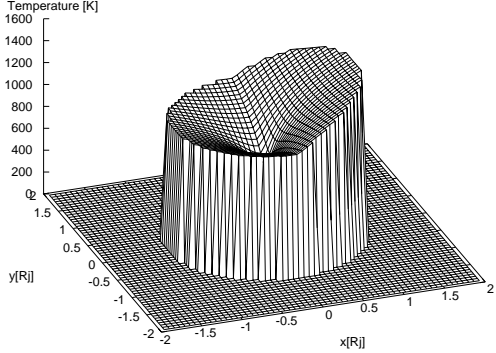}}
\caption{Example of a 2D projection map of the surface temperature of 
HD189733b calculated for $P_{r}=0.6, P_{a}=0.1$.
x-axis points to the star, z-axis to the rotation pole. One can clearly 
see the hot region on the day side of the planet facing the star.
The night side has a non zero temperature due to the efficient heat
redistribution ($P_{r}=0.6$). The temperature at the polar
region, $(x,y)=(0,0)$, drops significantly since the heat is redistributed
efficiently along parallels but much less effectively along the meridians 
($P_{a}=0.1$).
}
\label{f0}  
\end{figure}

\begin{figure}
\centerline{\includegraphics[width=6.cm,angle=-90,clip=]{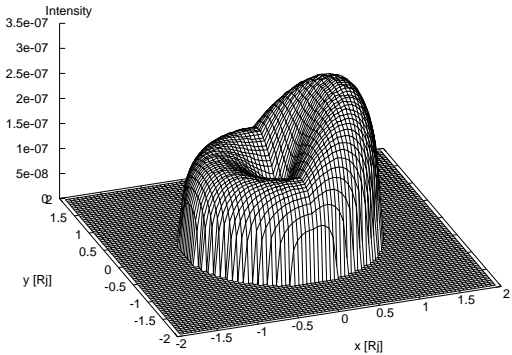}}
\caption{Example of a 2D projection image of HD189733b at 8 micron seen 
pole-on. 
The surface intensity is in $erg/s/Hz/cm^{2}/sterad$.
It corresponds to the temperate distribution from Fig.\ref{f0}.
The limb darkening ($u_{1}=0.6, u_{2}=0.2$) was applied to it.
Notice that hot regions are now actually dark due to the limb darkening.
}
\label{f0b}  
\end{figure}

First, let us introduce the qualitative distribution of the temperature
over the surface of HD189733b produced by our model.
Fig.\ref{f0} illustrates the behaviour of the temperature
for a relatively intense heat redistribution factor $P_{r}=0.6$
and quite small zonal temperature redistribution parameter $P_{a}=0.1$ 
(i.e. intense flows along parallels but not very intense heat flows
along meridians).
One can observe that this model produces hotter regions at 
the sub-stellar point and near the equator while cooler regions are
at the poles.
Fig.\ref{f0b} then illustrates the behaviour of the intensity.
This is what would an observer see looking on the planet pole-on. 
Limb darkening was applied to the model. One would see the dark polar 
regions, however, the hot sub-stellar point would not be the brightest 
since all hotter equatorial regions would be dim due to the limb 
darkening. Black body approximation was used to model the energy 
distribution.

\begin{figure}
\centerline{\includegraphics[width=5.5cm,angle=-90,clip=]{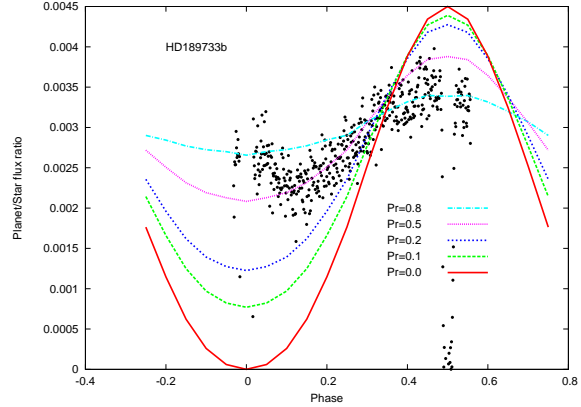}}
\caption{Theoretical light-curves of HD189733b compared with the observations
of \citet{knutson07} at 8 micron (black points).
Observe the strong dependence on the heat redistribution parameter, $P_{r}$,
which could reach several orders of magnitude at phase zero.
$A_{B}=0, P_{a}=1.0$, and zero limb darkening were assumed.
}
\label{f1}
\end{figure}

Now, we can proceed to a comparison between the observations of 
\citet{knutson07} and models produced by SHELLSPEC shown in Fig.\ref{f1}.
The synthetic light-curves are very sensitive to the heat redistribution
parameter $P_{r}=<0,1>$. Especially, at the phases close to the transit,
when the night side of the planet is seen, the flux can change
by the orders of magnitude. The best fit is  obtained for $P_{r}=0.5-0.8$,
which means that the surface heat redistribution is crucial and
quite effective. This finding is in agreement with the results of 
\citet{knutson07}, \citet{burrows08}, \citet{ms09}, and \citet{ca10}.

\begin{figure}
\centerline{\includegraphics[width=5.5cm,angle=-90,clip=]{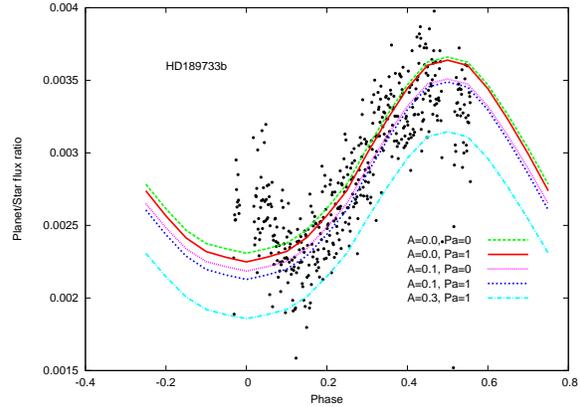}}
\caption{The effect of the Bond albedo and zonal temperature distribution
on the light-curve of transiting exoplanet HD189733b at 8 micron.
Higher Bond albedo reflects more light which is seen mainly at shorter
wavelength. Consequently, less energy is available to be absorbed and 
redistributed over the surface.
Temperatures are lower which means lower fluxes in the IR region.
If $P_{a}=0$ and the heat flows mainly along the parallels  
then (compared to the homogeneous flows with $P_{a}=1$) one would detect 
slightly more light at all phases, especially on the night side. 
Black points are the observations of \citet{knutson07}.
$P_{r}=0.6$ and zero limb darkening were assumed.
}
\label{f2}
\end{figure}

Albedo is an important parameter at all wavelengths.
The effect of the wavelength independent Bond albedo is illustrated 
in Fig.\ref{f2}. Our models indicate that Bond albedo of this planet
is relatively small of about 0.1 which is in agreement with \citet{ca10}.
Higher albedo reflects more light which could be observed
especially at the shorter wavelengths. This reflected light
might be relatively small in the IR region. However, higher albedo
reduces the amount of energy absorbed and redistributed over the surface.
Surface temperatures are lower which manifests in lower fluxes at longer
wavelengths. 

It is interesting to see that these kinds of light curves of transiting
exoplanets are not very sensitive to the zonal temperature redistribution 
parameter $P_{a}=<0,1>$ associated with the effectiveness of 
the homogeneous heat transfer versus zonal transfer. 
It is illustrated in Fig.\ref{f2} too.
If the heat flows mainly along the equator and not along the meridians
($P_{a}=0$) and we view the planet almost edge-on then 
we observe a slight increase of flux at all phases, especially 
on the night side (compared to the homogeneous temperature distribution,
$P_{a}=1$).

\begin{figure}
\centerline{\includegraphics[width=5.5cm,angle=-90,clip=]{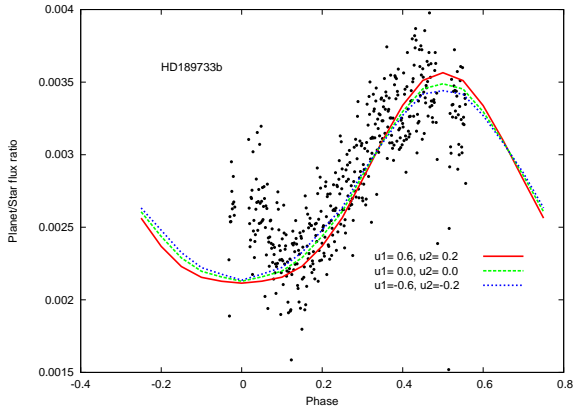}}
\caption{The effect of the limb darkening/brightening on the theoretical
light-curve of the transiting exoplanet HD189733b at 8 micron.  Limb
darkening has almost no effect at phase zero but much stronger effect
shortly before and after the transit and near the secondary eclipse.  This
is because when we observe the night side with almost constant temperature
distribution the darkening at the limb is compensated by the brightening at
the center to conserve the flux.  However, during the secondary eclipse the
limb darkening suppresses the radiation from the cold regions and amplifies
the radiation from the hot regions.  The net effect is that planet is
brighter at the phase 0.5.  The opposite happens shortly before and after
the transit.  The possible limb brightening would have the opposite
behaviour.  Black points are the observations of \citet{knutson07}.
$A_{B}=0.1, P_{r}=0.6, P_{a}=1.0$ were assumed.  
} 
\label{f3}
\end{figure}

It is also interesting to observe a moderate effect of 
the limb darkening on such light-curves of transiting exoplanets.
Limb darkening is manifested mainly shortly before and after the transit
and near the secondary eclipse. See Fig.\ref{f3} with the comments
and the description.
It might be interesting to point out that extrasolar planets might
show limb brightening at some wavelengths.
This depends on the temperature gradient at a particular depth probed
by a certain wavelength. If the temperature is decreasing one would observe
the limb darkening. If it is increasing one would observe limb brightening.
There are two main effects which could cause the temperature inversion.
(1) Presence of species high in the atmosphere effectively absorbing near 
the wavelength of maximum of the stellar energy redistribution. 
This would cause the temperature inversion often called a stratosphere.
(2) Heat redistribution between the day and the night side, especially 
if it occurs at deeper layers \citep{burrows08}.
This might significantly cool such layers on the day side producing
a drop in the temperature and associated temperature inversion.
This is the main motivation why we study also the effect of limb
brightening on the Fig.\ref{f3} and its effect is just the opposite
of the limb darkening. It might be worth mentioning that all the above
mentioned calculations took into account the proper inclination
of the orbit. Depending how close to edge-on the planet's orbit is, 
the observed light curve may vary by 5\% simply due to the observer's 
viewing angle \citep{ca08}.
 
\begin{figure}
\centerline{\includegraphics[width=5.5cm,angle=-90,clip=]{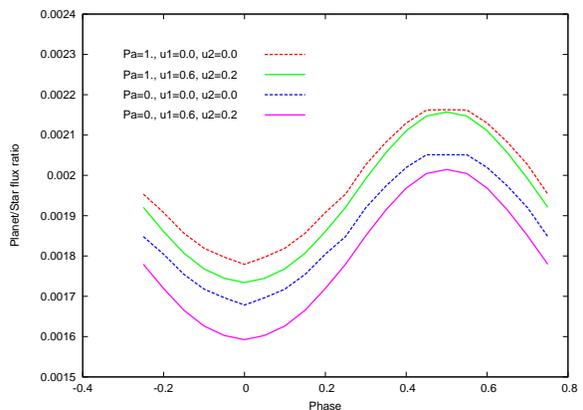}}
\caption{The effect of the zonal temperature redistribution 
($P_{a}$ parameter) and limb darkening on the light-curve of 
a hypothetical exoplanet seen at small inclination of 20 degrees.
Contrary to the case with homogeneous temperature distribution ($P_{a}=1$),
one would observe less flux from the models with the dominant
east-west heat circulation ($P_{a}=0$) since the polar regions visible
at this inclination are quite cool.
Limb darkening would dim the hotter equatorial regions even further
and reduce the observed flux, especially for models with the dominant
east-west heat circulation ($P_{a}=0$).
$A_{B}=0, P_{r}=0.5$ were assumed.
}
\label{f4}
\end{figure}

However, there are cases when the zonal temperature redistribution 
parameter, $P_{a}$, and limb darkening will be much more important.
Fig.\ref{f4} illustrates the situation for a hypothetical planet
at an inclination of 20 degrees. We used the parameters for the
non-transiting extrasolar planet HD179949b \citep{tinney01,cowan07}
to feed the code with real numbers. However, inclination, mass and radius
are not known for non-transiting exoplanets.
In this case the effects of the zonal temperature redistribution 
parameter and limb darkening are indeed important and are comparable to 
the amplitude of the light curve. 
Another point worth making is that there will be a strong degeneracy
between the radius, $P_{r}$, and inclination since all of them affect
the amplitude of the light-curve. Also there may be a degeneracy
between $P_{a}$, limb darkening and radius since all of them affect 
the mean level of the light-curve.
This justifies the complexity of our model with the heat redistribution
and zonal temperature redistribution, and limb darkening parameters.
Gravity darkening is also included but this will not be an issue 
for the extrasolar planets. If they are close to the star then 
their intrinsic radiation is negligible compared to the heavy external 
irradiation and if they are far from the star then their shape will 
be almost spherical (neglecting the rotation) and without the gravity
darkening.

\begin{figure}
\centerline{\includegraphics[width=5.5cm,angle=-90,clip=]{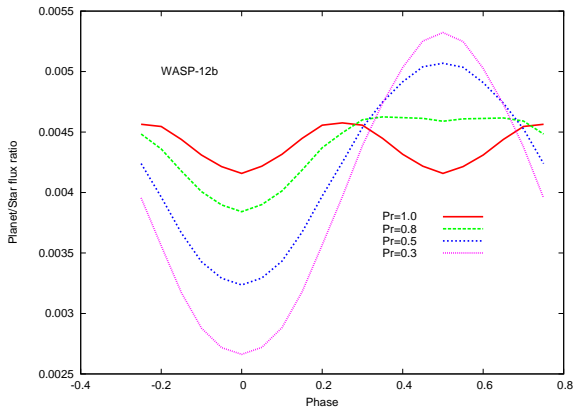}}
\caption{Illustration of the effect of the non-spherical shape on 
the light-curve of WASP-12b at 8 microns.
In case of effective heat redistribution ($P_{r}=1$),
one would observe a highly atypical double humped light-curve
caused by the Roche shape. This resembles the well known ellipsoidal 
variations in interacting binaries.
For lower values of $P_{r}$ and less effective heat redistribution
the light curve would acquire a more typical 'cosine' shape.
$A_{B}=0, P_{a}=1$ and zero limb darkening were assumed.
}
\label{f5}
\end{figure}

Finally, let us study how this Roche shape might affect the light-curve
of an extrasolar planet. This is illustrated Fig.\ref{f5} on the example
of WASP-12b at 8 micron. The light varies by about 10\% 
as predicted by \citet{li10}. However, the most striking 
thing is the double humped light-curve in case of effective heat 
redistribution $P_{r}=1$. This shape of the light-curve gradually 
changes to a typical single humped shape for less intense heat 
redistribution (lower values of $P_{r}$). The transition to a single 
humped light-curve is color dependent and for shorter wavelengths 
occurs at higher $P_{r}$.
These calculations assumed the proper inclination of the orbit,
zero albedo, zero limb darkening, and $P_{a}=1$.
Assuming $A_{B}=0, P_{r}=1, P_{a}=1, T_{old}=100K$ and Roche shape
we got the mean temperature of the planet of 2470 K which is in 
good agreement with 2516K obtained by \citet{hebb09} who assumed
the spherical shape.

\begin{figure}
\centerline{\includegraphics[width=5.5cm,angle=-90,clip=]{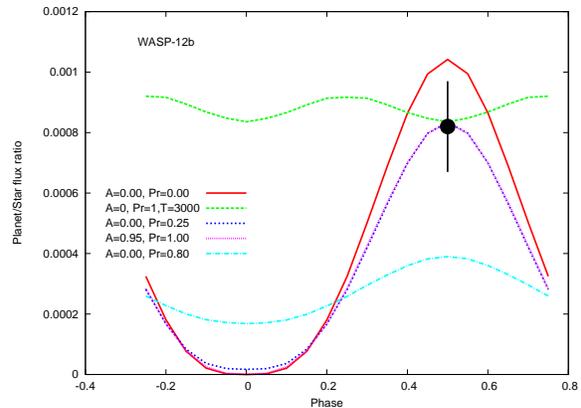}}
\caption{Light-curves of WASP-12b at 0.9 microns.
Observations of \citet{morales09}, black point, can be understood
within several very different models.
Low albedo and low heat redistribution model ($A_{B}=0, P_{r}=0.25$)
produces almost identical light-curve than high albedo models 
($A_{B}=0.95$, $P_{r}$ is not very important in this case).
Also a non-reflective model with homogeneous temperature distribution
of about 3000 K ($A_{B}=0, P_{r}=1, P_{a}=1$) can reproduce 
the observations. Roche shape and zero limb darkening were assumed.
}
\label{f5b}
\end{figure}

In the next step we calculated the light-curves of the planet
at much shorter wavelength (0.9 micron) and attempted to understand 
the secondary eclipse observations of \citet{morales09}.
This is illustrated in Fig.\ref{f5b}.
Notice that at this shorter wavelength the light-curves cover 
a considerable range of values during the secondary eclipse. 
This is contrary to the longer wavelengths when light-curves have higher
spred during the transit. There is a lot of degeneracy between 
the albedo and heat redistribution in this case.
For example the observation of \citet{morales09} can be reproduced
equally well by the model with $A_{B}=0, P_{r}=0.25$ with the mean
surface temperature 2470 K as well as by the model with
$A_{B}=0.95, P_{r}=1$ with the mean surface temperature of 1170K.
Note that heat redistribution is not very important nor well
constrained if the albedo is very high.
It is also possible to fit the observations assuming that 
the surface is not reflective and the temperature is homogeneous over 
the entire surface ($A_{B}=0, P_{r}=1,P_{a}=1$) and adopting
$T_{old}=2600K$. This corresponds to the mean temperature of the planet 
of about 3000K (in this case the mean temperature is identical to 
the effective temperature and to the brightness temperature of the body). 
This is much higher than a day side brightness temperature of 2660 K 
reported by \citet{morales09} but it is in perfect agreement with 
the value of 2997 K given by \citet{ca10}.
%??These authors assumed $A_{B}=0.3, P_{n}=0.5$ which
%does not make much sence since this combination of parameters it would
%result in night side temperatures well above the days side temperatures. 

\subsection{Spectra of the extrasolar planets}

I would like to point out that this paper does not deal with 
the calculations of the local atmosphere models and intensity emerging
from the local atmosphere models on the surface of the planet or star. 
However, in case grids of such models and spectra are available 
\citep{hb07,allard03} they can be used as an input to our model and 
composite spectra of the star and/or planet can be calculated from 
optional view point. In this respect, our code is not restricted 
to the black body approximation.

\begin{figure}
\centerline{\includegraphics[width=5.5cm,angle=-90,clip=]{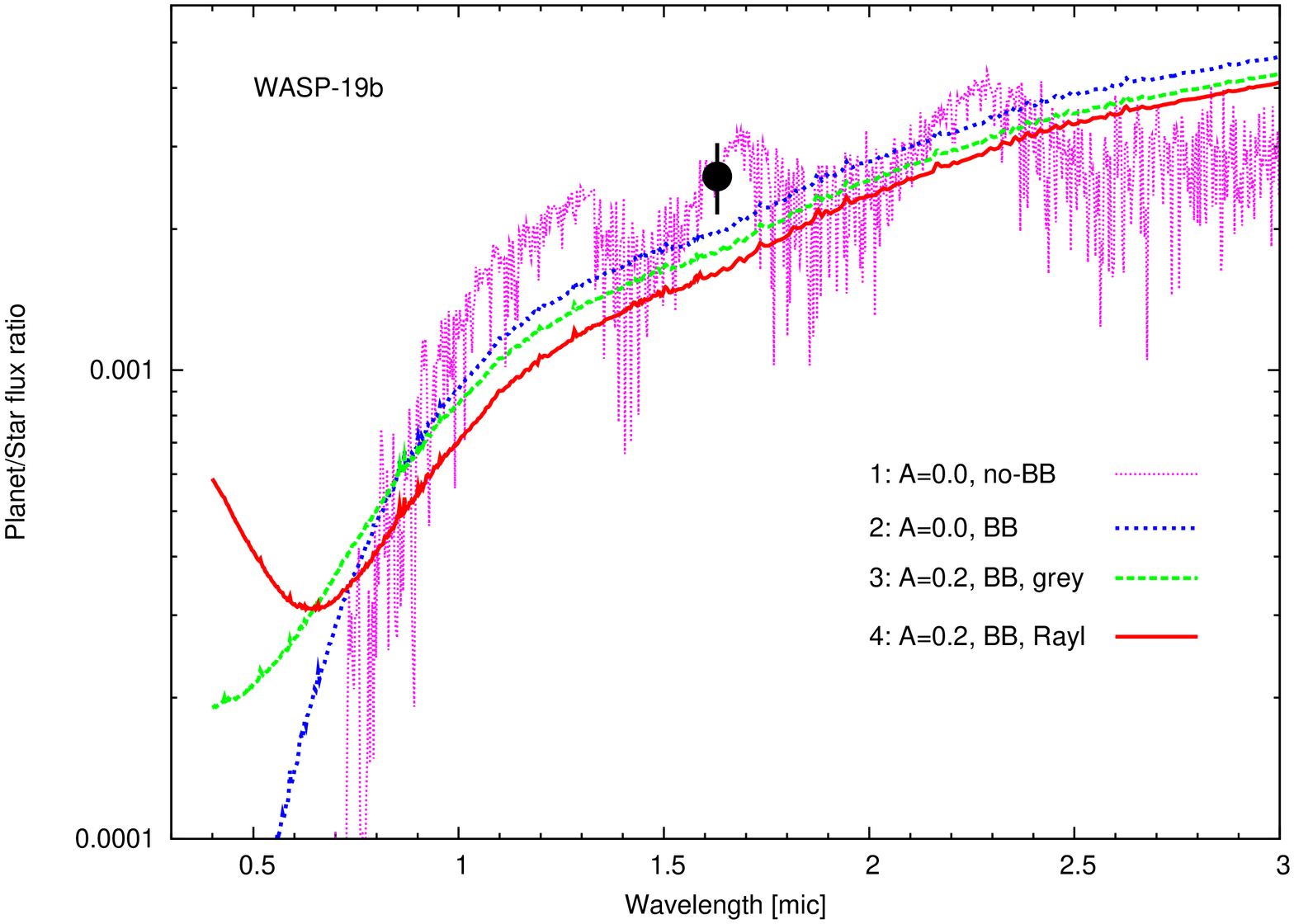}}
\caption{Planet to star flux ratio as a function of wavelength for WASP-19b.
Model (2) is a black-body model with zero albedo. 
Model (3) is a grey albedo model ($A_{B}=0.2$). 
Model (4) is a non-grey albedo model (Rayleigh scattering) and the same
Bond albedo, $A_{B}=0.2$.
Model (1) is a non black-body model based on non-irradiated atmospheres
\citep{hb07,allard03}.
Black symbol is a measurement of \citet{anderson10b} in the H-band.
Zero heat and homogeneous zonal temperature redistribution 
($P_{r}=0, P_{a}=1$) and zero limb darkening were assumed.
}
\label{f6}
\end{figure}

In this section we apply our model to WASP-19b and study its spectrum.
This planet was discovered by \citet{hebb10}.
The secondary eclipse was detected in the H-band by \citet{anderson10b}.
The authors found the emission from the planet very strong
and it presents an interesting puzzle for the models. 
This is a highly distorted exoplanet and we assumed a proper
shape according to the Roche potential, proper inclination,
zero heat redistribution, homogeneous zonal temperature redistribution, 
zero limb and gravity darkening and calculated spectra during 
the secondary eclipse.
It was also assumed that the star has a spectrum given by \citet{ck04}. 
As a first approximation for the planet I used the black body
approximation.
The comparison of the data with our model is in Fig.\ref{f6}.
Our black body model with zero albedo and zero heat redistribution
$P_{r}=0$ still under-predicts the radiation from the planet
but it is within $2\sigma$ from the measurements.

This figure also illustrates the effect of the albedo on the spectrum.  
For example, relatively small frequency independent (grey) albedo 
of about 0.2 will strongly increase the planet radiation at the shorter
wavelengths but decrease the fluxes from the planet at wavelengths longer
than 0.9 micron. This is because the non-zero albedo reduces the surface
temperature of the planet but reflects the stellar light at shorter
wavelengths. In the next step we assumed that the albedo is not grey 
but has the following color dependence
\begin {equation}
A_{\nu}=\frac{A_{0}(\lambda/\lambda_{0})^{\gamma}}
{(1-A_{0})+A_{0}(\lambda/\lambda_{0})^{\gamma}}.
\end {equation}
where $A_{0}$ is albedo at some reference wavelength $\lambda_{0}$.
We set $\gamma=-4$ to mimic the Rayleigh scattering with
with $\lambda^{-4}$ dependence and adjust $A_{0}$ to recover the same
Bond albedo (0.2) as in the previous case.
This case produces identical surface temperatures as in the grey
albedo case but radiation below 0.6 micron is considerably higher
than in the grey case, radiation above the 0.6 micron is lower
than in the grey case, and at still longer wavelength the relative
difference between the grey and color albedo calculations
gradually decreases.
Finally, I refrained from the black-body approximation for the planet
and used the grid of non-irradiated atmosphere models
of \citet{hb07} for $T_{eff}<1800K$ and \citet{allard03} (BT-Settl) 
for $T_{eff}>1800K$.
With this approach one can fit the observations of \citet{anderson10b}
reasonably well. The fit requires that the planet has low Bond albedo
and low heat redistribution. The stratosphere, if present, should not
have drastic impact on the spectrum since if the JHK band were reversed
and were in absorption (instead of emission) it might be difficult
to conserve the total flux and fit these observations simultaneously.

\section{Conclusions - Summary}

I argue that the reflection effect in the interacting binaries 
must be revisited in order to describe properly the radiation 
from the cool object irradiated by the hot object.
Subsequently, a new model for the reflection effect was proposed
and applied to an extreme case of an interacting binary - extrasolar
planet.

The new model introduces several free parameters. Some of them are well
known in the field of extrasolar planets.
(1) Bond albedo $A_{B}=<0,1>$ which controls how much energy is reflected 
and how much is converted into the heat. $A_{B}=1$ means that all impinging 
energy is reflected off the surface and nothing gets converted into the heat.
(2) Heat redistribution parameter $P_{r}=<0,1>$ which controls how much
heat from a certain point is redistributed to other places and how much
is re-radiated locally. $P_{r}=0$ means that all absorbed heat is re-radiated
locally and nothing gets transported to other places.
(3) Zonal temperature redistribution parameter $P_{a}=<0,1>$ 
is a measure of the effectiveness of the homogeneous heat redistribution
over the surface versus the zonal distribution. $P_{a}=1$ means that 
the heat is homogeneously redistributed over the surface while $P_{a}=0$
means zonal redstribution and that the heat flows only along the parallels.
Our model properly describes the shapes of the objects
by means of the Roche potential and takes into account gravity and limb
darkening. At the same time it takes into account the orbital revolution,
rotation and proper Doppler shifts in the scattered and thermal radiation.
%The code is freely available with the complete documentation??
%and example runs at http:\\www.ta3.sk/~budaj/shellspec

It is demonstrated on HD189733b that the light-curves of transiting 
extrasolar planets are mainly sensitive to the heat redistribution 
parameter $P_{r}$ and not so sensitive to the zonal temperature 
redistribution parameter $P_{a}$ or to the limb darkening. 
However, light-curves of planets seen at very low inclinations are 
very sensitive also to the zonal temperature redistribution parameter 
$P_{a}$ as well as to the limb darkening. This planet has low Bond albedo
and relatively intense heat redistribution.

The effect of non-spherical shape on the light-curve can also be important.
For highly distorted planets like WASP-12b this might cause 
a double humped curve with the amplitude of about 10\% superposed
on other types of variability. Observations of this planet at 0.9 micron
cannot constrain the Bond albedo or heat redistribution well.

The effect of the grey and non-grey albedo on the spectrum of the Wasp-19b
was studied as well. It was demonstrated that non-blackbody model
with fluxes given by non-irradiated atmosphere models \citep{hb07,allard03}
can fit the observations of \citet{anderson10b} reasonably well.
The planet has low Bond albedo and low heat redistribution.

We calculated the exact Roche shapes of all currently known
transiting exoplanets \citep{schneider95} and found out that the departures 
from spherical symmetry may vary significantly. 
Departures of the order of 1\% are common,
and can exceed about 8\% in most extreme cases like WASP-12b, WASP-19b, and
WASP-33b. About 8\% of transiting planets have departures more than 3\% 
(all have semi-major axes smaller than 0.03 AU).
Temperature redistribution over the surface of all these planets
was also calculated. The mean temperatures of the planets also vary 
considerably from 300 K to 2600 K. The extreme cases are
WASP-33b, WASP-12b, and WASP-18b with mean temperatures
of about 2600, 2430, and 2330 K, respectively.

\acknowledgments
I thank Nick Cowan for valuable comments on the manuscript and 
Heather Knutson for sharing a copy of her data.
The author acknowledges the support from the Marie Curie International
Reintegration grant FP7-200297 and partial support from 
the VEGA grants 2/0078/10, 2/0074/09.

\clearpage

\begin{table*}
\begin{center}
\caption{
Shapes of the transiting exoplanets.
Columns are:
$a$ -semi-major axis in [AU], $q$ -star/planet mass ratio,
$R_{sub}$  -planet radius at the sub-stellar point ($R_{sub}$,0,0),
$R_{back}$ -planet radius at the anti-stellar point,
$R_{pole}$ -planet radius at the rotation pole (0,0,$R_{pole}$),
$R_{side}$ -planet radius at the side point (0,$R_{side}$,0),
(assumed equal to the planet radius determined from the transit),
$R_{eff}$  -effective radius of the planet,
$rr=R_{sub}/R_{pole}$ -departure from the sphere,
$f_{i}=R_{sub}/L_{1x}$ -fill-in parameter of the Roche lobe,
$T$  -mean temperature of the planet in K.
Radii are in units of Jupiter radius. 
Coordinates of the center of mass of the planet are (0,0,0).
See the text for a more detailed information.
\label{T1}
}
%\scriptsize
\tiny
\begin{tabular}{llrllllllllll}
\tableline\tableline
name            &  a      &   q   &  Rsub   & Rback   & Rpole   & Rside   & Reff    & rr      &$f_{i}$& $T$  & Reference             \\ 
\tableline                                                                                     
GJ 1214b        & 0.01400 &  9186 & 0.24342 & 0.24340 & 0.24088 & 0.24150 & 0.24192 & 1.01053 & 0.253 &  551 & \citet{charbonneau09} \\ 
WASP-19b        & 0.01640 &   865 & 1.43896 & 1.42960 & 1.28052 & 1.31000 & 1.33680 & 1.12373 & 0.590 & 1914 & \citet{hebb10}        \\ 
CoRoT-7b        & 0.01720 & 64503 & 0.15108 & 0.15107 & 0.14965 & 0.15000 & 0.15024 & 1.00952 & 0.244 & 1760 & \citet{leger09}       \\ 
WASP-18b        & 0.02047 &   128 & 1.16966 & 1.16949 & 1.16349 & 1.16500 & 1.16601 & 1.00531 & 0.208 & 2329 & \citet{southworth09a} \\ 
OGLE-TR-56b     & 0.02250 &   949 & 1.34392 & 1.34211 & 1.28736 & 1.30000 & 1.30950 & 1.04393 & 0.414 & 1967 & \citet{pont07}        \\ 
TrES-3          & 0.02260 &   504 & 1.31652 & 1.31569 & 1.28839 & 1.29500 & 1.29967 & 1.02183 & 0.329 & 1606 & \citet{odonovan07}    \\ 
WASP-12b        & 0.02290 &  1002 & 1.98765 & 1.97291 & 1.74650 & 1.79000 & 1.83079 & 1.13808 & 0.613 & 2407 & \citet{hebb09}        \\ 
OGLE-TR-113b    & 0.02290 &   618 & 1.10252 & 1.10213 & 1.08607 & 1.09000 & 1.09273 & 1.01515 & 0.290 & 1114 & \citet{gillon06}      \\ 
WASP-4b         & 0.02300 &   840 & 1.46767 & 1.46538 & 1.40130 & 1.41600 & 1.42715 & 1.04736 & 0.425 & 1815 & \citet{southworth09b} \\ 
CoRoT-1b        & 0.02540 &   966 & 1.54395 & 1.54167 & 1.47462 & 1.49000 & 1.50166 & 1.04702 & 0.424 & 1835 & \citet{barge08}       \\ 
WASP-33b        & 0.02560 &  1410 & 1.66033 & 1.65542 & 1.53414 & 1.56000 & 1.58143 & 1.08225 & 0.512 & 2582 & \citet{cameron10}     \\ 
WASP-5b         & 0.02729 &   653 & 1.18130 & 1.18101 & 1.16773 & 1.17100 & 1.17325 & 1.01162 & 0.266 & 1737 & \citet{southworth09c} \\ 
CoRoT-2b        & 0.02810 &   306 & 1.47584 & 1.47548 & 1.46154 & 1.46500 & 1.46736 & 1.00979 & 0.253 & 1495 & \citet{alonso08}      \\ 
GJ 436b         & 0.02872 &  6574 & 0.43970 & 0.43968 & 0.43744 & 0.43800 & 0.43838 & 1.00516 & 0.200 &  695 & \citet{bean08}        \\ 
SWEEPS-11       & 0.03000 &   118 & 1.13119 & 1.13116 & 1.12961 & 1.13000 & 1.13026 & 1.00140 & 0.134 & 1959 & \citet{sahu06}        \\ 
OGLE-TR-132b    & 0.03060 &  1157 & 1.19343 & 1.19309 & 1.17577 & 1.18000 & 1.18294 & 1.01503 & 0.288 & 1925 & \citet{gillon07}      \\ 
HD 189733b      & 0.03099 &   741 & 1.14505 & 1.14488 & 1.13573 & 1.13800 & 1.13955 & 1.00821 & 0.236 & 1178 & \citet{bouchy05}      \\ 
WASP-2b         & 0.03138 &   962 & 1.02260 & 1.02249 & 1.01519 & 1.01700 & 1.01823 & 1.00730 & 0.227 & 1258 & \citet{daemgen09}     \\ 
WASP-3b         & 0.03170 &   737 & 1.32162 & 1.32131 & 1.30630 & 1.31000 & 1.31254 & 1.01172 & 0.266 & 1930 & \citet{gibson08}      \\ 
TrES-2          & 0.03556 &   856 & 1.28042 & 1.28022 & 1.26929 & 1.27200 & 1.27385 & 1.00876 & 0.241 & 1455 & \citet{daemgen09}     \\ 
WASP-24b        & 0.03590 &  1145 & 1.11018 & 1.11006 & 1.10200 & 1.10400 & 1.10536 & 1.00742 & 0.228 & 1611 & \citet{street10}      \\ 
XO-2b           & 0.03690 &  1800 & 0.97839 & 0.97830 & 0.97126 & 0.97300 & 0.97419 & 1.00735 & 0.226 & 1281 & \citet{burke07}       \\ 
WASP-14b        & 0.03700 &   178 & 1.26047 & 1.26044 & 1.25851 & 1.25900 & 1.25932 & 1.00156 & 0.138 & 1802 & \citet{joshi08}       \\ 
WASP-10b        & 0.03710 &   243 & 1.08107 & 1.08105 & 1.07965 & 1.08000 & 1.08024 & 1.00132 & 0.130 & 1008 & \citet{johnson09}     \\ 
HAT-P-7b        & 0.03790 &   855 & 1.43186 & 1.43160 & 1.41753 & 1.42100 & 1.42338 & 1.01011 & 0.253 & 2075 & \citet{welsh10}       \\ 
WASP-1b         & 0.03820 &  1459 & 1.37324 & 1.37289 & 1.35319 & 1.35800 & 1.36134 & 1.01482 & 0.287 & 1749 & \citet{cameron07}     \\ 
HAT-P-12b       & 0.03840 &  3623 & 0.96819 & 0.96804 & 0.95607 & 0.95900 & 0.96102 & 1.01267 & 0.271 &  930 & \citet{hartman09}     \\ 
HAT-P-3b        & 0.03894 &  1636 & 0.89290 & 0.89285 & 0.88905 & 0.89000 & 0.89064 & 1.00432 & 0.190 & 1119 & \citet{torres07}      \\ 
TrES-1          & 0.03930 &  1493 & 1.08664 & 1.08654 & 1.07917 & 1.08100 & 1.08224 & 1.00692 & 0.222 & 1050 & \citet{alonso04}      \\ 
WASP-26b        & 0.04000 &  1150 & 1.32922 & 1.32902 & 1.31704 & 1.32000 & 1.32202 & 1.00924 & 0.245 & 1615 & \citet{smalley10}     \\ 
HAT-P-5b        & 0.04075 &  1146 & 1.26718 & 1.26704 & 1.25768 & 1.26000 & 1.26158 & 1.00755 & 0.229 & 1496 & \citet{bakos07a}      \\ 
OGLE-TR-10b     & 0.04162 &  1961 & 1.27165 & 1.27142 & 1.25629 & 1.26000 & 1.26256 & 1.01223 & 0.268 & 1238 & \citet{pont07}        \\ 
WASP-6b         & 0.04210 &  2456 & 1.23659 & 1.23635 & 1.22000 & 1.22400 & 1.22677 & 1.01360 & 0.278 & 1353 & \citet{gillon09}      \\ 
WASP-16b        & 0.04210 &  1251 & 1.01088 & 1.01084 & 1.00706 & 1.00800 & 1.00864 & 1.00380 & 0.182 & 1235 & \citet{lister09}      \\ 
HAT-P-13b       & 0.04260 &  1503 & 1.28881 & 1.28863 & 1.27717 & 1.28000 & 1.28194 & 1.00911 & 0.243 & 1600 & \citet{bakos09a}      \\ 
HD 149026b      & 0.04313 &  3792 & 0.65543 & 0.65542 & 0.65353 & 0.65400 & 0.65432 & 1.00291 & 0.165 & 1700 & \citet{sato05}        \\ 
HAT-P-10b       & 0.04390 &  1867 & 1.04940 & 1.04933 & 1.04357 & 1.04500 & 1.04597 & 1.00558 & 0.206 & 1004 & \citet{bakos09b}      \\ 
HAT-P-4b        & 0.04460 &  1940 & 1.27961 & 1.27944 & 1.26692 & 1.27000 & 1.27211 & 1.01002 & 0.251 & 1641 & \citet{kovacs07}      \\ 
XO-3b           & 0.04540 &   107 & 1.21742 & 1.21741 & 1.21686 & 1.21700 & 1.21709 & 1.00046 & 0.092 & 1662 & \citet{johns08}       \\ 
WASP-28b        & 0.04550 &  1243 & 1.12346 & 1.12340 & 1.11887 & 1.12000 & 1.12076 & 1.00410 & 0.187 & 1375 & \citet{west10}        \\ 
Kepler-4b       & 0.04560 & 16634 & 0.35747 & 0.35747 & 0.35684 & 0.35700 & 0.35710 & 1.00175 & 0.139 & 1570 & \citet{borucki10}     \\ 
WASP-29b        & 0.04560 &  3479 & 0.74182 & 0.74181 & 0.73940 & 0.74000 & 0.74040 & 1.00328 & 0.172 &  970 & \citet{hellier10}     \\ 
Kepler-6b       & 0.04567 &  1892 & 1.33329 & 1.33310 & 1.31970 & 1.32300 & 1.32526 & 1.01030 & 0.253 & 1461 & \citet{dunham10}      \\ 
Lupus-TR-3b     & 0.04640 &  1124 & 0.89117 & 0.89115 & 0.88961 & 0.89000 & 0.89026 & 1.00175 & 0.140 &  987 & \citet{weldrake08}    \\ 
WASP-22b        & 0.04680 &  2057 & 1.12528 & 1.12520 & 1.11828 & 1.12000 & 1.12116 & 1.00626 & 0.214 & 1384 & \citet{maxted10}      \\ 
OGLE-TR-111b    & 0.04700 &  1620 & 1.07037 & 1.07032 & 1.06590 & 1.06700 & 1.06774 & 1.00419 & 0.188 &  987 & \citet{pont04}        \\ 
HD 209458b      & 0.04707 &  1544 & 1.32755 & 1.32742 & 1.31756 & 1.32000 & 1.32166 & 1.00759 & 0.229 & 1375 & \citet{henry00}       \\ 
WASP-25b        & 0.04740 &  1805 & 1.26718 & 1.26705 & 1.25768 & 1.26000 & 1.26158 & 1.00755 & 0.228 & 1207 & \citet{enoch10}       \\ 
Kepler-8b       & 0.04830 &  2106 & 1.43187 & 1.43162 & 1.41490 & 1.41900 & 1.42183 & 1.01200 & 0.266 & 1616 & \citet{jenkins10}     \\ 
XO-5b           & 0.04870 &   855 & 1.09073 & 1.09070 & 1.08843 & 1.08900 & 1.08938 & 1.00211 & 0.150 & 1207 & \citet{burke08}       \\ 
HAT-P-8b        & 0.04870 &   882 & 1.50647 & 1.50635 & 1.49790 & 1.50000 & 1.50142 & 1.00573 & 0.209 & 1658 & \citet{latham09}      \\ 
XO-1b           & 0.04880 &  1163 & 1.18727 & 1.18722 & 1.18293 & 1.18400 & 1.18472 & 1.00367 & 0.180 & 1208 & \citet{cullough06}    \\ 
CoRoT-5b        & 0.04947 &  2242 & 1.39965 & 1.39943 & 1.38428 & 1.38800 & 1.39056 & 1.01110 & 0.259 & 1400 & \citet{rauer09}       \\ 
WASP-15b        & 0.04990 &  2280 & 1.44095 & 1.44071 & 1.42387 & 1.42800 & 1.43085 & 1.01200 & 0.266 & 1607 & \citet{west09}        \\ 
Kepler-5b       & 0.05064 &   680 & 1.43466 & 1.43459 & 1.42980 & 1.43100 & 1.43180 & 1.00340 & 0.176 & 1758 & \citet{borucki10b}    \\ 
TrES-4          & 0.05091 &  1577 & 1.82048 & 1.81998 & 1.79224 & 1.79900 & 1.80371 & 1.01576 & 0.292 & 1705 & \citet{daemgen09}     \\ 
OGLE-TR-211b    & 0.05100 &  1352 & 1.36583 & 1.36573 & 1.35810 & 1.36000 & 1.36128 & 1.00569 & 0.208 & 1683 & \citet{udalski08}     \\ 
OGLE-TR-182b    & 0.05100 &  1182 & 1.13241 & 1.13238 & 1.12921 & 1.13000 & 1.13053 & 1.00284 & 0.165 & 1315 & \citet{pont08}        \\ 
WASP-17b        & 0.05100 &  2564 & 1.77093 & 1.77022 & 1.73048 & 1.74000 & 1.74678 & 1.02337 & 0.333 & 1594 & \citet{anderson10}    \\ 
WASP-21b        & 0.05200 &  3526 & 1.07550 & 1.07543 & 1.06821 & 1.07000 & 1.07121 & 1.00682 & 0.220 & 1229 & \citet{bouchy10}      \\ 
HAT-P-6b        & 0.05235 &  1278 & 1.33465 & 1.33457 & 1.32848 & 1.33000 & 1.33102 & 1.00464 & 0.194 & 1628 & \citet{noyes08}       \\ 
WASP-13b        & 0.05270 &  2937 & 1.21722 & 1.21711 & 1.20767 & 1.21000 & 1.21159 & 1.00791 & 0.231 & 1439 & \citet{skillen09}     \\ 
HAT-P-9b        & 0.05300 &  1718 & 1.40744 & 1.40731 & 1.39759 & 1.40000 & 1.40164 & 1.00705 & 0.223 & 1487 & \citet{shporer09}     \\ 
HAT-P-11b       & 0.05300 & 10473 & 0.45248 & 0.45248 & 0.45184 & 0.45200 & 0.45211 & 1.00142 & 0.130 &  844 & \citet{bakos10}       \\ 
SWEEPS-04       & 0.05500 &   341 & 0.81015 & 0.81014 & 0.80995 & 0.81000 & 0.81003 & 1.00024 & 0.073 & 1348 & \citet{sahu06}        \\ 
HAT-P-1b        & 0.05530 &  2264 & 1.23003 & 1.22996 & 1.22336 & 1.22500 & 1.22611 & 1.00545 & 0.205 & 1259 & \citet{bakos07b}      \\ 
XO-4b           & 0.05550 &   803 & 1.34252 & 1.34248 & 1.33917 & 1.34000 & 1.34055 & 1.00250 & 0.159 & 1414 & \citet{cullough08}    \\ 
CoRoT-3b        & 0.05700 &    66 & 1.01006 & 1.01006 & 1.00998 & 1.01000 & 1.01001 & 1.00008 & 0.052 & 1656 & \citet{deleuil08}     \\ 
HAT-P-14b       & 0.06060 &   650 & 1.15084 & 1.15083 & 1.14972 & 1.15000 & 1.15019 & 1.00098 & 0.116 & 1525 & \citet{torres10}      \\ 
WASP-7b         & 0.06180 &  1396 & 0.91568 & 0.91568 & 0.91477 & 0.91500 & 0.91515 & 1.00099 & 0.116 & 1344 & \citet{hellier09}     \\ 
Kepler-7b       & 0.06224 &  3258 & 1.48887 & 1.48871 & 1.47451 & 1.47800 & 1.48040 & 1.00974 & 0.248 & 1514 & \citet{latham10}      \\ 
HAT-P-2b        & 0.06878 &   156 & 1.15714 & 1.15714 & 1.15695 & 1.15700 & 1.15703 & 1.00016 & 0.065 & 1442 & \citet{pal09}         \\ 
WASP-8b         & 0.07930 &   485 & 1.17030 & 1.17030 & 1.16990 & 1.17000 & 1.17007 & 1.00034 & 0.082 &  912 & \citet{smith09}       \\ 
CoRoT-6b        & 0.08550 &   373 & 1.16618 & 1.16618 & 1.16594 & 1.16600 & 1.16604 & 1.00021 & 0.070 &  990 & \citet{fridlund10}    \\ 
CoRoT-4b        & 0.09000 &  1600 & 1.19072 & 1.19072 & 1.18976 & 1.19000 & 1.19016 & 1.00081 & 0.108 & 1039 & \citet{moutou08}      \\ 
HD 17156b       & 0.16230 &   404 & 1.02302 & 1.02302 & 1.02299 & 1.02300 & 1.02300 & 1.00002 & 0.033 &  852 & \citet{winn09}        \\ 
CoRoT-9b        & 0.40700 &  1234 & 1.05000 & 1.05000 & 1.05000 & 1.05000 & 1.05000 & 1.00000 & 0.019 &  401 & \citet{deeg10}        \\ 
HD 80606b       & 0.44900 &   239 & 0.92100 & 0.92100 & 0.92100 & 0.92100 & 0.92100 & 1.00000 & 0.009 &  365 & \citet{fossey09}      \\ 

\tableline
\end{tabular}
\normalsize
\end{center}
\end{table*}

\clearpage

\end{document}